\documentclass[%
 reprint,
 amsmath,amssymb,
 aps,
 prx,
]{revtex4-2}

\usepackage{graphicx}
\usepackage{dcolumn}
\usepackage{bm}
\usepackage{xcolor}
\usepackage{multirow}
\usepackage{soul}
\usepackage{comment}

\usepackage{tikz}
\usetikzlibrary{calc, angles, quotes}
\usepackage{hyperref}

\begin{document}

\preprint{APS/123-QED}

\title{{Origin of Persistent Boundary Motion in Confined Active Matter}}% Force line breaks with \\
%Orientational dynamics of an ABP at a curved interface
\author{Elsa Baby}
\affiliation{Department of Physics, Indian Institute of Technology Palakkad, India.}%Lines break automatically or can be forced with \\
\author{Manoj Gopalakrishnan}
\email{manojgopal@iitpkd.ac.in}
\affiliation{Department of Physics, Indian Institute of Technology Palakkad, India.}%Lines break automatically or can be forced with \\
\author{Vishwas V. Vasisht}%
\email{vishwas@iitpkd.ac.in}
\affiliation{%
 Department of Physics, Indian Institute of Technology Palakkad, India.%\textbackslash\textbackslash
}%


\date{\today}

\begin{abstract}
\noindent Active matter systems under confinement display persistent surface motion and a strong boundary affinity. However, despite extensive studies of their positional dynamics, much less attention has been given to the corresponding orientational behavior. Here, using molecular simulations of an active Brownian particle confined within a hard circular boundary and the Fokker–Planck equation, we show that the positional distribution of the particle is directly coupled to orientational fluctuations, as characterized by the conditional orientational distribution. Confinement generates two preferred tangential orientational states connected by stochastic flipping pathways: rapid boundary-localized switching and slower bulk-mediated excursions. Further, the positional distribution exhibits a nontrivial power-law decay with distance from the boundary that is closely linked to curvature-induced bistable orientational states and the variance of the associated conditional distribution. The mean waiting time between flips exhibits power-law dependence on the confinement strength. Our results establish that the interplay between orientational fluctuations, bistability, positional accumulation, and stochastic switching governs the observed dynamics of active particles under confinement, providing a framework for understanding transport, exploration, and escape processes in confined active systems.

%Active Brownian particles within a confinement display persistent surface motion and a strong affinity to the boundary. However, despite extensive studies of their positional dynamics, much less attention has been given to the corresponding orientational behavior. Here we show, using molecular simulations and the Fokker–Planck equation, the boundary motion of a single active Brownian particle confined within a circular domain is governed by stochastic flipping between two preferred tangential orientational states. We uncover two distinct pathways connecting these states: rapid boundary-localized flips and slower bulk-mediated excursions. The mean waiting time between flips exhibit power-law decay with the confinement strength, defined as the ratio of persistence length to confinement radius.  The positional distribution exhibits a nontrivial power-law decay as a function of distance from the boundary which is shown to be closely linked to curvature-induced bistable orientational states and the variance in the associated conditional distribution. The interplay between confinement, orientational fluctuations, bistability and spatial organization governs the stochastic switching dynamics underlying anomalous transport, exploration, and escape processes in confined active systems.
\end{abstract}
\maketitle

%\tableofcontents
\section{\label{secIntro}Introduction}
Active matter consists of systems whose constituents continuously consume energy to generate directed motion, thereby driving the system far from thermal equilibrium \cite{ramaswamy2017active,fodor2016far}. Over the past decade, active particle systems have become a major area of research across physics, biology, and materials science. These systems span a wide range of scales, from molecular motors and bacterial suspensions to synthetic colloidal swimmers and active granular media \cite{marchetti2013hydrodynamics, ramaswamy2010mechanics, bechinger2016active}. Their nonequilibrium nature gives rise to a variety of emergent phenomena, including collective motion, anomalous transport, pattern formation, and motility-induced phase separation \cite{vicsek2012collective,Sujit_RSC_2023,czirok2000collective,cates2015motility}. Active particles often operate in complex and confined environments where geometric constraints strongly influence their dynamics. In biological systems, microorganisms navigate through porous media, tissues, and microfluidic channels \cite{bhattacharjee2019porous,toley2012tissue,overberg2024microchanel}, while in synthetic systems active dopants are increasingly used to modify the mechanical and transport properties of passive materials \cite{ni2013activedopants,kummel2015activedopants,sharma_NPhys_2025}. The persistent nature of active motion leads to nontrivial interactions with boundaries, obstacles, and interfaces, substantially altering transport and exploration dynamics. Understanding the microscopic dynamics of active particles in such confined environments is therefore essential for connecting single-particle behavior to the emergent properties of active systems.\\
\indent A prominent consequence of confinement is the strong affinity of active particles toward surfaces, where even simple geometries can generate persistent sliding motion and long residence times near boundaries. Such behavior has been observed experimentally in bacteria, sperm cells, and synthetic Janus swimmers near planar and curved surfaces \cite{berke2008hydrodynamic, denissenko2012hspermwall, das2015januswall}. While early studies often attributed boundary accumulation to hydrodynamic interactions \cite{lauga2006swimming, berke2008hydrodynamic, tuson2013bacteria}, later experiments and simulations showed that even active particles without hydrodynamic interactions can strongly accumulate near boundaries due to the interplay of confinement, self-propulsion, and rotational diffusion \cite{drescher2011fluid, elgeti2009self, elgeti2013wall, elgeti2015physics}. Previous studies further reported non-Boltzmann steady states, strong boundary localization, and positional distributions exhibiting exponential-to-power-law crossovers under confinement \cite{elgeti2013wall, takatori2016acoustic, dauchot2019dynamics, schmidt2021non, buttinoni2022active}. Motivated by these observations, in a recent work we investigated the positional dynamics of a single bacterium confined within a vesicle using experiments, molecular simulations, and a Fokker--Planck description based on an active Brownian particle model \cite{vincent2025curvature}. We showed that the positional distribution crosses over from rapid exponential decay near the boundary to a slower decay before approaching a uniform bulk limit, with the associated length scales governed by translational and rotational diffusion. Existing studies of confined active matter, including our earlier work, have therefore focused primarily on positional observables while largely integrating over particle orientation.\\
\indent Orientational dynamics is expected to play a central role in the accumulation of active particles near confining boundaries. Early theoretical studies showed that positional accumulation near hard walls is accompanied by strong orientational anisotropy, with particles preferentially aligning parallel to the boundary \cite{elgeti2013wall}. Subsequent works demonstrated that confinement can induce localized polar order in otherwise isotropic active systems, with the ordering becoming stronger in concave geometries \cite{duzgun2018active}. Related studies on active Brownian particles confined within harmonic \cite{malakar2020steady, nakul2023stationary} and superharmonic traps \cite{samui2025active} further revealed strongly biased and bimodal orientational distributions associated with persistent orbiting motion and boundary-localized steady states. Despite these advances, most existing studies have focused primarily on average orientational order in stationary states, while the role of orientational fluctuations and stochastic switching dynamics underlying persistent boundary motion remains largely unexplored. Stochastic switching between opposite sliding states provides a natural mechanism governing persistent motion near curved boundaries and the associated transport and residence dynamics. This makes orientational fluctuations particularly relevant for exploration, escape, and geometric sorting \cite{bechinger2016active, araujo2023steering} in confined active systems.\\
\indent Motivated by these considerations, in this work we investigate the coupled positional and orientational dynamics of a single active Brownian particle confined within a hard circular boundary using molecular simulations and Fokker--Planck framework. To isolate the minimal mechanism underlying boundary-induced orientational dynamics, we focus on the high-Péclet limit with vanishing translational diffusion, where the dynamics are governed solely by the ratio of persistence length to confinement radius. We show that even this minimal confinement generates a positional distribution exhibiting a nontrivial power-law decay with an exponent 2/3 near the boundary before crossing over to a uniform bulk profile. The corresponding conditional orientational distribution reveals curvature-induced bistable states associated with persistent sliding along the boundary, together with a pronounced asymmetry possibly arising from the hard-wall constraint. Its peak value and variance further exhibit power-law scaling with distance from the boundary, characterized by exponents $1/3$ and $2/3$, respectively. Most importantly, the interplay between confinement, orientational fluctuations, and bistability gives rise to two distinct stochastic flipping pathways, namely rapid boundary-localized flips and slower bulk-mediated excursions, whose mean waiting time between flips exhibit power-law scaling with confinement strength.\\
\indent The paper is organized as follows. In Sec.~II, we introduce the model and describe the simulation and analytical methods employed in this work. The main results on the positional distribution, orientational distribution, and their mutual coupling are presented in Secs.~III, IV, and V, respectively. The stochastic orientational flipping dynamics are discussed in Sec.~VI, followed by the conclusions in Sec.~VII.
\vspace{-3mm}
\section{Model and Methods}
\noindent {\bf Simulation details: } We model the confined active particle as an active Brownian particle (ABP) in a two-dimensional circular domain with radius $R$ and purely reflecting boundaries. The particle dynamics are obtained by numerically integrating the coupled Langevin equations of motion \cite{vincent2025curvature}
\begin{eqnarray}
m\ddot{\mathbf r} &=& -\gamma_t\left(\dot{\mathbf r} - u_0\hat{\mathbf u}\right) +{\boldsymbol{\nu}_t}, \label{1}\\
\dot{\theta} &=& \nu_r , \label{2}
\end{eqnarray}
where $\mathbf r$ and $\theta$ denote the particle position and orientation, respectively, $\gamma_t$ is the translational viscous drag coefficient, and $u_0\hat{\mathbf u}$ is the self-propulsion velocity with orientation vector $\hat{\mathbf u}=(\cos\theta,\sin\theta)$. The translational thermal noise ${\boldsymbol{\nu}}_t(t)$ has zero mean and autocorrelation $ \langle {\boldsymbol{\nu}}_t(t)\cdot{\boldsymbol{\nu}}_t(t^{\prime})\rangle
=4k_B T\gamma_t\delta (t-t^{\prime})$,
with the associated translational diffusion coefficient $D_t=k_B T/\gamma_t$. To isolate activity–confinement effects from translational thermal fluctuations \cite{vincent2025curvature}, we consider the limit $D_t/(R^2D_r)\to0$, such that translational diffusion is neglected by setting $D_t=0$ and hence ${\boldsymbol{\nu}}_t(t)=0$. The rotational noise $\nu_r(t)$ is Gaussian white noise with autocorrelation
$\langle \nu_r(t)\nu_r(t') \rangle = 2D_r\delta(t-t')$, where $D_r$ is the rotational diffusion coefficient. The corresponding persistence time and persistence length are given by $\tau_P=D_r^{-1}$ and $l_P=u_0/D_r$, respectively.

\begin{figure}[htbp]
    \centering  \includegraphics[width=0.7\columnwidth]{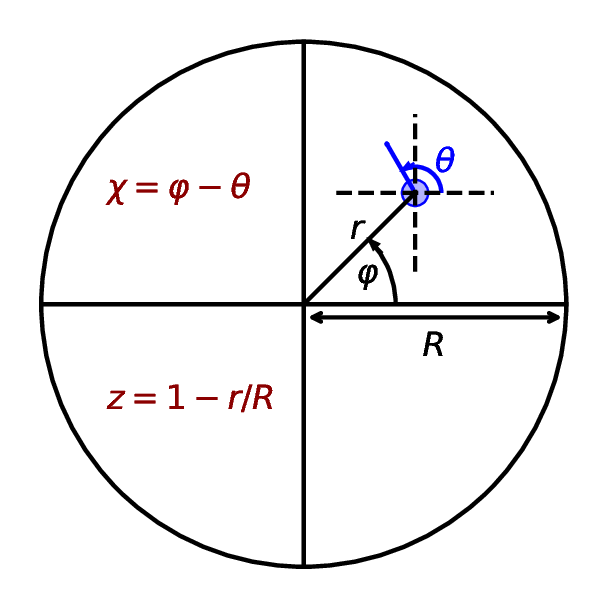}   
    \caption{Schematic illustration of an active Brownian particle confined within a circular domain of radius $R$. The particle position is described by the radial coordinate $r$ and polar angle $\varphi$, while $\theta$ denotes the propulsion orientation. The relative orientational variable is defined as $\chi=\theta-\varphi$. To characterize the boundary dynamics, we introduce the scaled radial coordinate $z=1-r/R$ and relative orientation $\chi$. Here $z=0$ corresponds to the boundary, $z=1$ to the center of the confining domain, $\chi=0$ corresponding to outward radial motion and $\chi=\pm\pi/2$ to tangential sliding along the boundary.}
    \label{fig1}
\end{figure}

Unless otherwise stated, simulations are performed in reduced units with parameters $m=1$, $u_0=0.025$, integration time step $\Delta t=0.001$ and $\gamma_t=10$, driving the system into an overdamped regime. The rotational diffusion coefficient is varied over the range $10^{-3}\leq D_r \leq 10^{-1}$, while the confinement radius is varied between $4 \leq R \leq 100$. Statistical averages are obtained from simulations extending up to approximately $10^8$ $\tau_P$ steps and over 10--50 independent realizations. To characterize the boundary dynamics, we introduce the scaled radial coordinate $z=1-r/R$ and the relative orientational variable $\chi=\theta-\varphi$ (Fig.~\ref{fig1}). Here, $z=0$ corresponds to the boundary and $z=1$ to the center of the confining domain, while $\chi$ measures the propulsion direction relative to the local radial direction, with $\chi=0$ corresponding to outward radial motion and $\chi=\pm\pi/2$ to tangential sliding along the boundary. The combined effects of persistence and confinement are quantified by the dimensionless confinement strength, defined as %\cite{vincent2025curvature}
\begin{eqnarray}
\tilde{u} &=& \frac{u_0 \tau_p}{R} = \frac{u_0}{R D_r}.
\label{3}
\end{eqnarray}

\noindent
{\bf Fokker-Planck equation: } The complete Fokker--Planck equation governing the joint probability density $\Pi(r,\chi;t)$ of the radial position $r$ and the relative orientational variable $\chi$ for a confined ABP in the overdamped regime \cite{malakar2020steady, vincent2025curvature} is given by 
\begin{eqnarray}
\frac{\partial \Pi}{\partial t}
&=&
D_t\nabla_r^2 \Pi
+\left(D_r+\frac{D_t}{r^2}\right)\frac{\partial^2 \Pi}{\partial \chi^2}
+\frac{u_0}{r}\sin\chi\frac{\partial \Pi}{\partial \chi}
\nonumber\\
&&
-\frac{\partial}{\partial r}
\bigg(
\left[u_0\cos\chi-N(r,\chi)\right]\Pi
\bigg),
\label{t1}
\end{eqnarray}
where $\nabla_r^2=\partial^2/\partial r^2+r^{-1}\partial/\partial r$ is the radial part of the two-dimensional Laplacian operator and $N(r,\chi)$ denotes the normal reaction force exerted by the confining boundary.  The four different terms in the RHS of Eq.~\eqref{t1} respectively represent translational diffusion, effective rotational diffusion, curvature-induced angular drift, and the radial propulsion current modified by the boundary reaction force.  For most of the results of this work, we focus on the stationary state and therefore set $\partial \Pi/\partial t =0$. After putting $D_t$ = 0 and in terms of the scaled radial coordinate $z$ (Fig.~\ref{fig1}), the dimensionless stationary probability density $P(z,\chi)=\lim\limits_{t\to\infty} R^2\Pi(r,\chi;t)$ satisfies the simpler equation
\begin{equation}
\frac{\partial}{\partial z}\bigg[\big(\cos\chi-n(z,\chi)\big)P\bigg]
+
\frac{1}{\tilde u}\frac{\partial^2 P}{\partial \chi^2}
+
\frac{\sin\chi}{1-z}\frac{\partial P}{\partial \chi}
=0,
\label{t2}
\end{equation}
where $n(z,\chi)=N(r,\chi)/u_0$ is the scaled normal reaction force. The distribution is symmetric under the transformation $\chi\to -\chi$, reflecting the equivalence of clockwise and anticlockwise motion.

The positional probability density is obtained by marginalizing the joint distribution over the orientational degree of freedom,
\begin{equation}
\phi(z)=\int_{-\pi}^{\pi}P(z,\chi)d\chi,
\label{t3}
\end{equation}
while the orientational statistics at a fixed position is characterized by the conditional orientational distribution
\begin{equation}
F(\chi|z)=\frac{P(z,\chi)}{\phi(z)}.
\label{t4}
\end{equation}
 Conditional orientational moments are correspondingly defined as
\begin{equation}
\langle \cos^n\chi\rangle_z =\int_{-\pi}^{\pi} F(\chi|z)\cos^n\chi d\chi.
\label{eq:COSN}
\end{equation}
{\bf Boundary conditions: } To formulate the boundary conditions, it is convenient to define the radial distribution function
\begin{equation}
f(z,\chi)=(1-z)P(z,\chi),
\label{eq:radial}
\end{equation}
together with the radial and orientational probability currents
\begin{eqnarray}
j_z(z,\chi) &=& -\big(\cos\chi-n(z,\chi)\big)f,
\label{eq:r-current}\\
j_\chi(z,\chi) &=& -\frac{1}{\tilde u}\frac{\partial f}{\partial\chi}-\frac{f\sin\chi}{1-z}
\label{eq:current}
\end{eqnarray}
The stationary Fokker-Planck equation may then be rewritten in the conservation form,
\begin{equation}
\frac{\partial j_z}{\partial z}+\frac{\partial j_\chi}{\partial\chi}=0,
\label{eq:FPE-f}
\end{equation}
which expresses probability conservation in the combined positional-orientational phase space.

Near the boundary, two distinct boundary regions emerge naturally. The first is a boundary-contact region, where the normal reaction force is finite and directly opposes outward propulsion. Beyond this lies the near boundary region, where the normal force vanishes but confinement-induced orientational effects remain significant. To capture this structure, we approximate the scaled normal reaction force as
\begin{equation}
n(z,\chi)\simeq
\cos\chi \ \Theta(z_0-z),
\label{eq:normal}
\end{equation}
where $\Theta(z)$ is the Heaviside step function and $z_0$ defines the outer edge of the boundary-contact region. 
The reflecting nature of the boundary requires the radial probability current to vanish at the boundary,
\begin{equation}
j_z(0,\chi)=0.
\label{eq:boundary-current}
\end{equation}
Assuming this condition remains valid throughout the boundary-contact region, Eq.~\eqref{eq:FPE-f} reduces to
\begin{equation}
\frac{\partial^2 f}{\partial \chi^2}+\tilde u\frac{\partial}{\partial\chi}\left(\sin\chi f(0, \chi)\right)=0,
\label{eq:boundary-distribution1}
\end{equation}
with solution
\begin{equation}
F(\chi|0)\propto e^{\tilde u\cos\chi},
\label{eq:boundary-distribution2}
\end{equation}
Equation~\eqref{eq:boundary-distribution2} therefore predicts a unimodal $F(\chi | z)$ at the boundary-contact region, which is also corroborated by the numerical simulations (see Fig.\ref{fig3} (b-d), top inset). Such a behavior corresponds to preferential outward radial alignment at the boundary. The crossover from the boundary-contact region to near boundary is observed in simulations as a sharp drop in the positional distribution (see dotted line in Fig.\eqref{fig2} main panel). For $z>z_0$ and $n(z,\chi)=0$, the vanishing radial probability current suppresses the outward propulsion component, and the dynamics is dominated by motion parallel to the boundary (see Appendix~A for further details). 

Next, we present our results for the positional distribution $\phi(z)$ at varying $\tilde{u}$ before discussing the conditional orientational distribution. Unless otherwise stated, symbols denote simulation results, while solid lines represent analytical predictions in all subsequent sections.

\section{Positional Distribution} \label{sec-pos}
\begin{figure}[htbp]
\centering
\includegraphics[width=1.0\columnwidth]{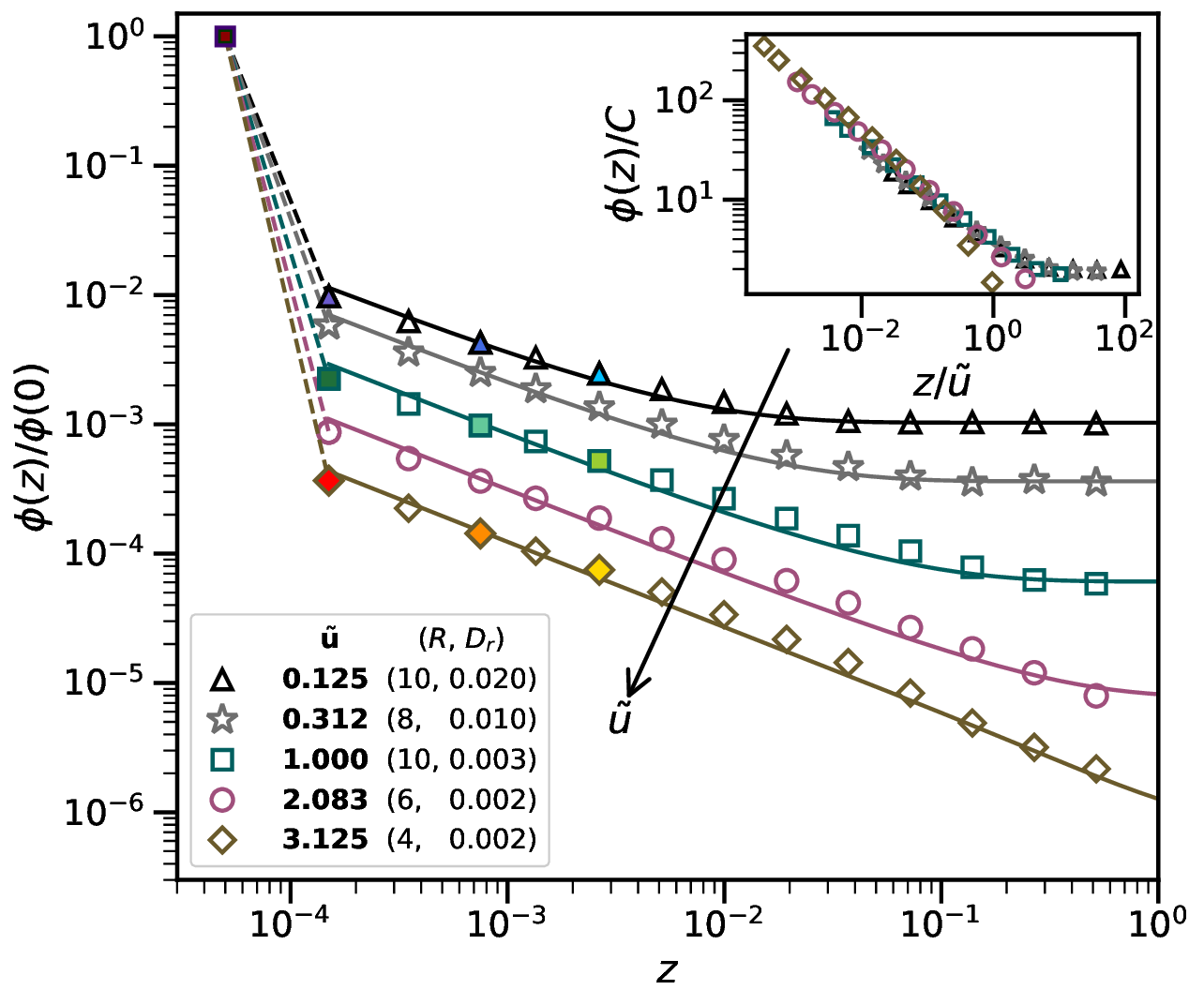}
\caption{{\bf Positional Distribution } Scaled positional distribution $\phi(z)/\phi(0)$ for different confinement strengths $\tilde{u}=u_0/(RD_r)$, where $u_0$ is the fixed propulsion velocity. Here $\phi(0)$ denotes the peak value of the normalized distribution. Dotted lines indicate the transition from the boundary-contact region. Symbols represent simulation data and solid lines correspond to analytical fits from Eq.~\eqref{4}. The filled symbols highlight the $z$ values where conditional orientational distribution are computed. (Inset) Collapse of the positional distribution obtained by rescaling $\phi(z)$ by $C$ and $z$ by $\tilde{u}$, where $C$ denotes the normalization constant obtained numerically from simulations and associated with Eq.~\eqref{4}.}
\label{fig2}
\end{figure}

In Fig.~\ref{fig2} we show the positional probability density $\phi(z)$ of the confined ABP for different values of the confinement strength $\tilde{u}$. Beyond the boundary-contact region, the distribution exhibits a slow power-law decay with increasing $z$. In our previous work \cite{vincent2025curvature}, for non-zero $D_t$, the positional distribution showed an initial rapid exponential decay crossing over to a slower decay before approaching a uniform bulk limit. In the absence of translational diffusion, only the slower decay survives, consistent with the behavior observed here. Similar exponential-to-power-law crossovers have also been reported for active particles confined between parallel plates \cite{elgeti2013wall}. Our simulations further show that increasing the confinement strength $\tilde{u}$ leads to a progressively steeper decay of $\phi(z)$. To understand both the power-law decay and its dependence on $\tilde{u}$, we now turn to the Fokker--Planck equation, where the stationary solution yields the following exact expression for $\phi(z)$:
\begin{equation}
\phi(z)\propto \frac{1}{\langle\langle \cos^2\chi\rangle\rangle_z}\exp\left(-\int_{0}^{z}\frac{d z^{\prime}}{(1-z^{\prime})\langle\langle \cos^2\chi\rangle\rangle_{z^{\prime}}}\right).
\label{eq:lambda2-phi*}
\end{equation}
The above solution already indicates a strong coupling between the positional distribution and the orientational fluctuations through the variance of $\cos\chi$. To understand this connection further, we compute the conditional orientational distribution $F(\chi|z)$ at selected values of $z$ (represented by the filled colored symbols in Fig.~\ref{fig2}). As discussed later, $F(\chi|z)$ will be used to derive analytical expressions for the variance and the positional distribution defined in Eqs.~\ref{eq:COSN} and \ref{t3}, respectively, which are then compared with the simulation results. We first present a detailed investigation of $F(\chi|z)$ in the next section.

\section{Orientational Distribution}\label{sec-orientation}
\begin{figure*}[htbp]
    \centering 
    \includegraphics[width=0.95\textwidth]{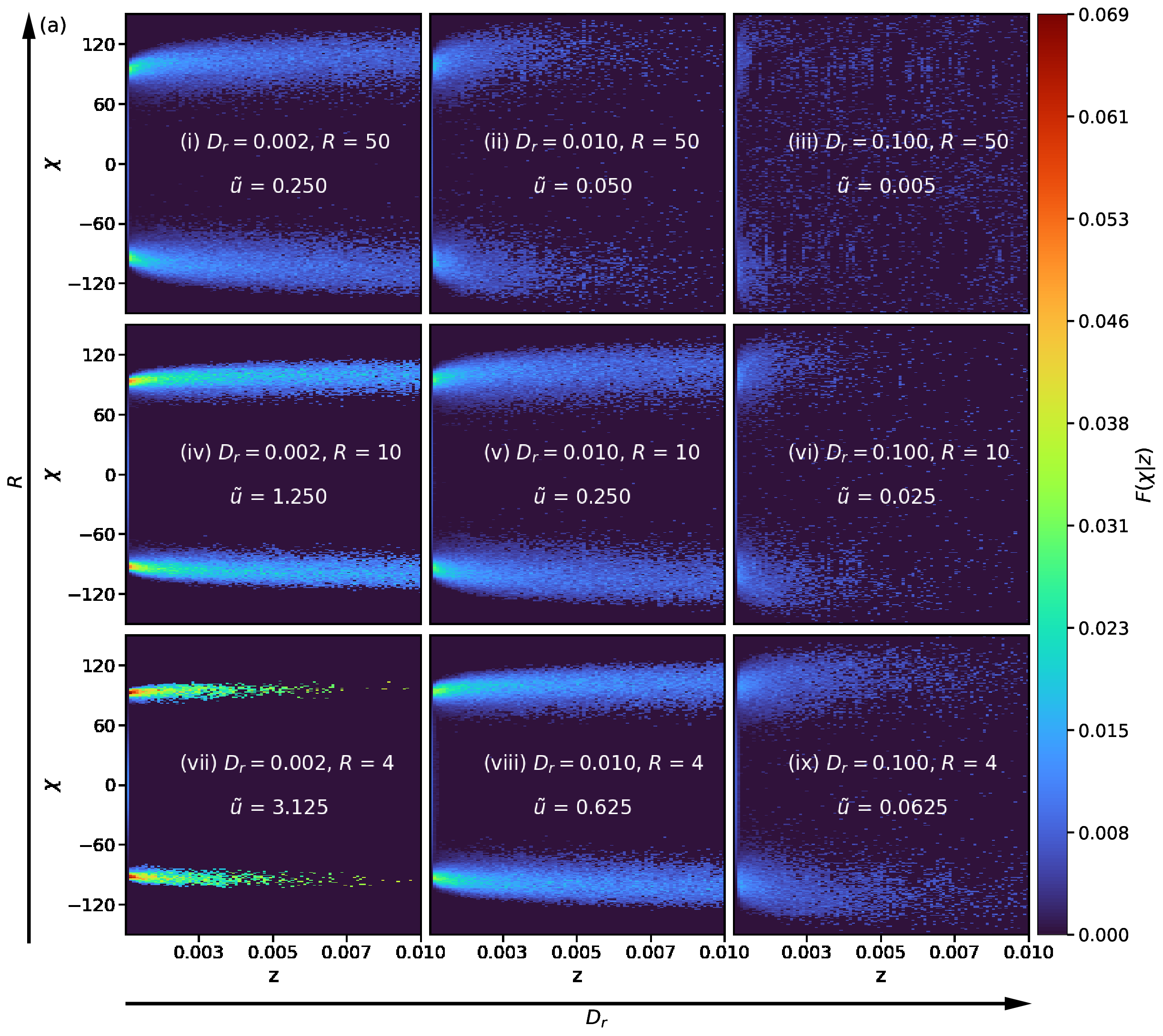}   
    \vspace{-2mm} 
    \includegraphics[width=0.95\textwidth]{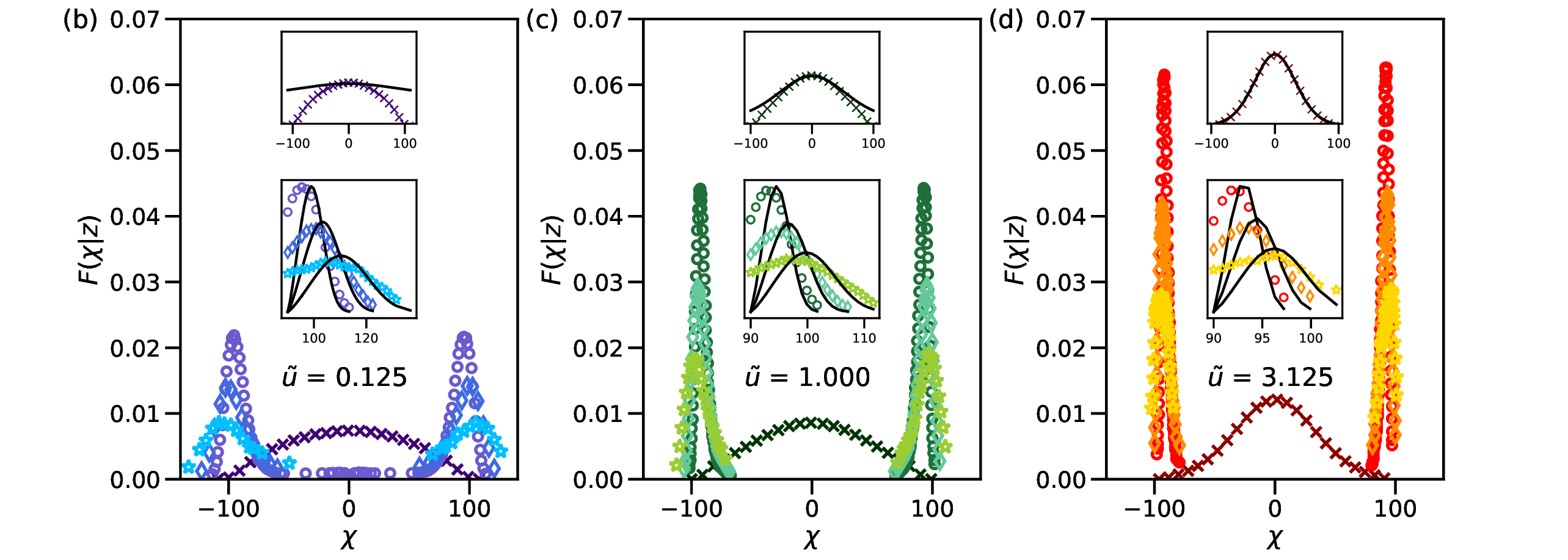}   
	\caption{{\bf Conditional Orientational Distribution} (a) Contour plots of the conditional orientational distribution $F(\chi|z)$ for different combinations of confinement radii $R$ and rotational diffusion coefficients $D_r$, with the corresponding confinement strengths $\tilde{u}=u_0/(RD_r)$ indicated in each panel, where $u_0$ is the fixed propulsion velocity. (b)--(d) Conditional orientational distributions $F(\chi|z)$ at selected values of $z$, with colors matching the corresponding filled symbols in Fig.~\ref{fig2}, for increasing confinement strengths $\tilde{u}=0.25$, $1.00$, and $3.125$. (Insets in b--d) Analytical fits to the simulation data are shown, where the top inset corresponds to the unimodal regime (Eq.~\eqref{eq:boundary-distribution2}) and the bottom insets correspond to the bimodal regime (Eq.~\eqref{7}).} %\red{Where is a, b, c, d in the figures? Put it outside the figure and make this consistency in each figure}
\label{fig3} 
\end{figure*}

\noindent {\bf Conditional orientational distribution $F(\chi | z)$: } In Fig.~\ref{fig3} we present our results for the conditional orientational distribution $F(\chi|z)$. We show the contour plots of $F(\chi|z)$ in the $(\chi,z)$ plane (Fig.~\ref{fig3}(a)) for increasing $D_r$ (left to right) and increasing confinement radius $R$ (top to bottom), with the corresponding confinement strengths $\tilde{u}$ indicated in the legend. We observe two distinct jets corresponding to symmetric peaks around $\chi=\pm\pi/2$, indicating the emergence of preferred tangential states associated with clockwise and anticlockwise sliding along the boundary. As the distance from the boundary increases, these jets broaden and weaken, reflecting the gradual evolution of the conditional orientational distribution away from the boundary. We further find that for large confinement strengths $\tilde{u}$ (left column), $F(\chi|z)$ remains sharp and well defined, whereas for smaller $\tilde{u}$ (right column) the distribution becomes increasingly diffuse, indicating the weakening of preferential orientational ordering.

The same behavior is shown more clearly in Fig.~\ref{fig3}(b--d), where $F(\chi|z)$ is plotted at selected values of $z$ (with the coloring scheme matching the corresponding datasets in Fig.~\ref{fig2}) for three representative confinement strengths $\tilde{u}=0.125$, $1.0$, and $3.125$. For a fixed $\tilde{u}$, the conditional distribution exhibits a pronounced bimodal structure peaked around $\chi=\pm\pi/2$. Moving away from the boundary, the peaks shift away from the purely tangential directions and the distribution progressively broadens, indicating a gradual weakening of confinement-induced orientational ordering. The variance of $F(\chi|z)$ at a fixed z, decreases with increase in $\tilde{u}$.

Before discussing the analytical predictions obtained from the Fokker--Planck formalism and comparing them with the simulation results, we briefly re-emphasize the behavior of $F(\chi|z)$ at the boundary-contact region. The distribution becomes unimodal and centered around $\chi=0$, corresponding to predominantly outward radial orientations. In this regime, the ABP remains pinned to the boundary while retaining a finite orientational freedom that permits limited motion along the boundary. This behavior naturally emerges as the limiting case discussed in Sec.~II.

We now turn to the analytical description of the orientational dynamics very close to the boundary. The emergence of the tangential states, associated with clockwise and anticlockwise sliding along the boundary, follows from the vanishing radial current condition at the boundary, $j_z(z_0,\chi)=0$, which reduces to $\cos\chi f(z_0,\chi)=0$ in Eq. \eqref{eq:r-current}. This yields the following conditional distribution describing the orientational structure in the immediate vicinity of the boundary.
\begin{equation}
F(\chi|z_0) = \frac{1}{2}\left[\delta(\chi-\pi/2)+\delta(\chi+\pi/2)\right].
\label{t5}
\end{equation}

While Eq.~\eqref{t5} predicts perfectly tangential states exactly at the boundary, the distribution acquires a finite width as particles move away from the boundary, as observed in the simulations (see Fig. \ref{fig3}). To characterize this broadening, we perform an asymptotic analysis of the stationary Fokker–Planck equation in Eq.~\ref{eq:FPE-f} close to $\chi=\pm\pi/2$. Using Eq.~\eqref{t5} as the boundary condition, we obtain the following expression for the conditional orientational distribution, which captures the finite-width bimodal structure observed in the simulations (see Appendix~B and C for details).

\begin{eqnarray}
F(\chi|z)&\propto&z^{-\tfrac{3}{4}}\left(\chi-\tfrac{\pi}{2}\right)^{\tfrac{5}{4}}\exp\left[-\tfrac{\tilde{u}}{9z}\left(\chi-\tfrac{\pi}{2}
\right)^3\right],
\label{7}
\end{eqnarray}

The simulation data (see the bottom insets of Fig.~\ref{fig3}(b--d)) are well described by the analytical form given in Eq.~\eqref{7}. Because Eq.~\eqref{7} is derived from an asymptotic expansion about the tangential states \(\chi=\pm\pi/2\), its validity is expected to extend from the tangential orientations into the right tail of the distribution. Accordingly, the theory accurately reproduces both the peak and its decay for \(\chi\geq\pi/2\), while deviations appear for \(\chi<\pi/2\), where the simulations retain a broader orientational weight.

Having characterized the full conditional distribution and its orientational structure, we now turn to the conditional orientational moments of $\cos\chi$.

\begin{figure}[htbp]
    \centering  \includegraphics[width=1.0\columnwidth]{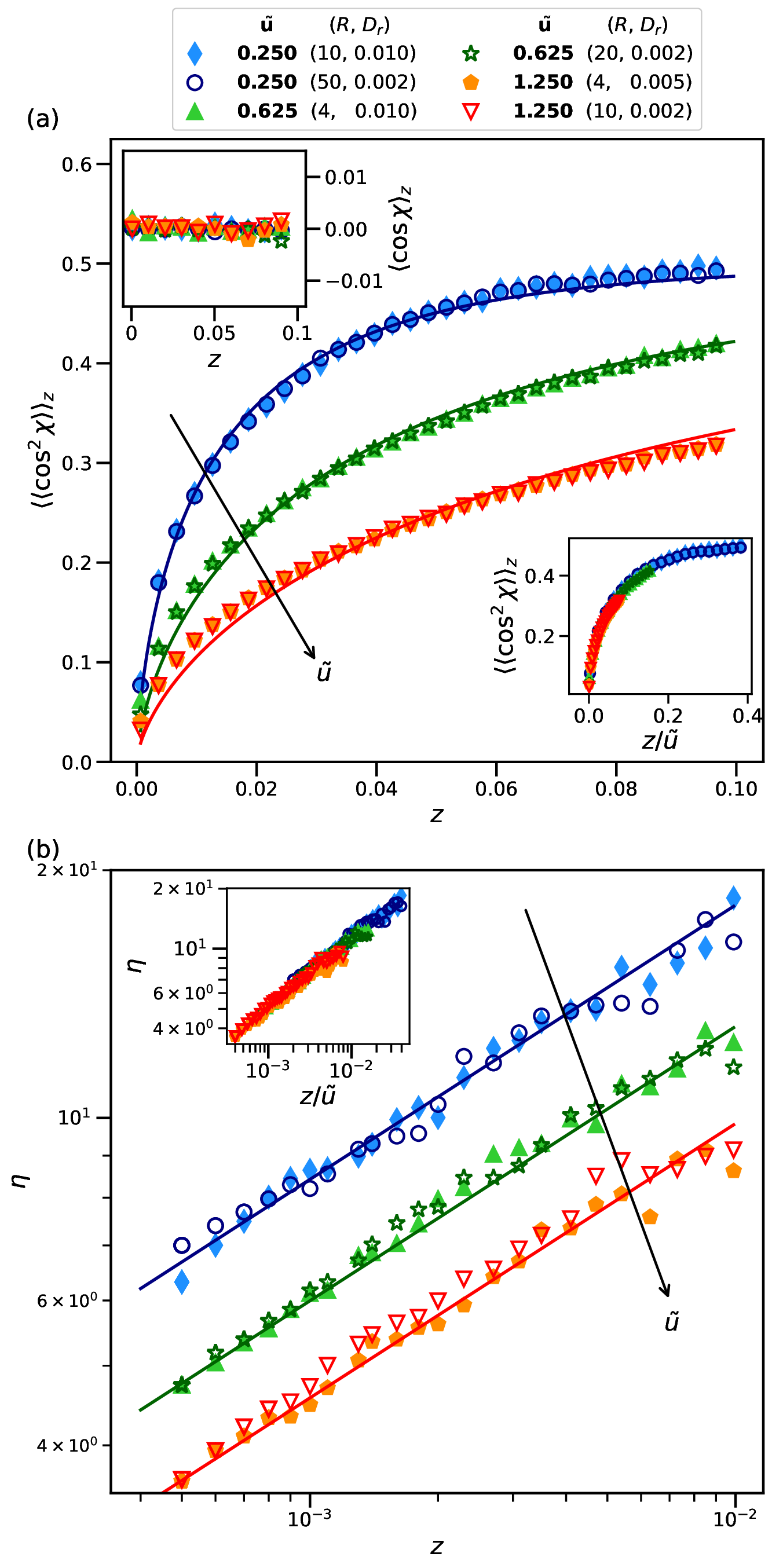}  
	\caption{{\bf Conditional Variance and Modal Angle} (a) Conditional variance of $\cos\chi$ as a function of $z$ for different confinement strengths $\tilde{u}=u_0/(RD_r)$ obtained from different combinations of confinement radii $R$ and rotational diffusion coefficients $D_r$, where $u_0$ is the fixed propulsion velocity. For each $\tilde{u}$, data corresponding to two distinct $(R,D_r)$ combinations (open and filled symbols) collapse onto a single curve. Insets show the first moment as a function of $z$ and the collapse of the variance for different $\tilde{u}$ under rescaling of $z$ by $\tilde{u}$. (b) Modal angle of the conditional distribution measured relative to the tangential orientation, defined as $\eta\equiv|\chi_{\mathrm p}-\pi/2|$, as a function of $z$ for the same parameter sets. Inset shows the collapse of the modal angle for different $\tilde{u}$ under rescaling of $z$ by $\tilde{u}$.}
 \label{fig4}
\end{figure}

\noindent {\bf Conditional Moments and Modes: }To quantify the orientational fluctuations associated with the bimodal regime, we analyze the conditional moments and modal angles of the distribution. The first moment satisfies $\langle\cos\chi\rangle_z=0$ owing to the symmetry between the two tangential states, making higher-order moments the leading quantities governing the confined dynamics (see ~\ref{fig4}(a) top inset). The simulation data for the conditional variance $\langle\langle\cos^2\chi\rangle\rangle_z$ are shown in Fig.~\ref{fig4}(a) as a function of the distance from the boundary. The variance remains strongly suppressed close to the wall, reflecting the sharply localized orientational distribution around the tangential states $\chi=\pm\pi/2$. As $z$ increases, the variance grows monotonically and eventually saturates near $1/2$, signaling the gradual recovery of isotropic orientational behaviour.

The behavior of the variance follows directly from the theoretical description of the conditional orientational distribution developed in the previous section. Combining the exact relation in Eq.~\eqref{eq:lambda2-phi*} with the asymptotic positional distribution close to the boundary, $\phi(z)\propto (z/\tilde u)^{-2/3}$, yields a direct description of the orientational fluctuations. Close to the boundary one obtains
\begin{equation}
\langle\langle\cos^2\chi\rangle\rangle_z\propto\left(\frac{z}{\tilde u}\right)^{2/3}.
\end{equation}
Motivated by this asymptotic scaling at $z\to0$ together with the isotropic limit $\langle\langle\cos^2\chi\rangle\rangle_z\to1/2$, we propose the following interpolating form (see Appendices~A and ~C for details):
\begin{equation}
\langle\langle \cos^2\chi\rangle\rangle_z =\frac{1}{2}
\left[1-\exp\left(-\left(\frac{z}{\tilde{u}}\right)^{\frac{2}{3}}\right)\right],\qquad(z_0\leq z\ll1)
\label{8}
\end{equation}
The above expression in  Eq.~\eqref{8} fits the simulation data for the conditional variance in Fig.~\ref{fig4}(a) remarkably well across the full range of confinement strengths. It accurately reproduces both the growth of fluctuations away from the boundary and the crossover to the isotropic saturation limit, indicating a universal scaling behavior governed by the confinement parameter $\tilde u$. The inset of Fig.~\ref{fig4}(a) shows the collapsed data obtained by scaling the $x$-axis with $\tilde u$.

The modal angle of the conditional distribution provides a complementary characterization of the orientational dynamics. As particles move away from the boundary, the peaks of the relative orientational distribution shift away from the perfectly tangential orientations at $\pm \pi/2$, indicating a gradual departure from boundary-guided motion. To quantify this shift, we define
\begin{eqnarray}
\eta &\equiv& |\chi_p-\pi/2|,
\label{10}
\end{eqnarray}
where $\chi_p$ denotes the peak orientation angle of $F(\chi|z)$. As shown in Fig.~\ref{fig4}(b), the simulation data exhibit a systematic increase of $\eta$ with distance from the wall. The theoretical prediction obtained from the extremum condition of Eq.~\eqref{7} captures the data well and yields the scaling
\begin{eqnarray}
\eta &\propto& z^{1/3}.
\label{11}
\end{eqnarray}
The growth of $\eta$ reflects the progressive weakening of confinement-induced orientational order. Near the boundary, strong confinement favors nearly tangential trajectories, while farther away enhanced orientational fluctuations allow particles to explore broader angular configurations. Similar to the conditional variance, the modal shift also exhibits a universal scaling behavior governed by the confinement parameter $\tilde u$, as shown by the data collapse in the inset of Fig.~\ref{fig4}(b).

\section{Position--Orientation Coupling}
Having established the complete description of the orientational distribution and its fluctuations, we now return to Eq.~\eqref{eq:lambda2-phi*} discussed in Sec.~III, which predicts the coupling between the positional distribution and orientational fluctuations. Substituting the conditional variance $\langle\langle\cos^2\chi\rangle\rangle_z$ from Eq.~\eqref{8} into Eq.~\eqref{eq:lambda2-phi*} yields a closed analytical form for the positional boundary layer. Introducing the transformed variable $\tau=-\ln(1-z)$, which maps the radial distance from the wall onto an effective evolution parameter, Eq.~\eqref{eq:lambda2-phi*} becomes
\begin{equation}
\phi(z) \propto 
\frac{1}{1-\exp\left(-\left[\frac{\tau}{\tilde{u}}\right]^{2/3}\right)}
\exp\left\{-3\tilde{u}\,
\gamma\left(\frac{3}{2},
\left[\frac{\tau}{\tilde{u}}\right]^{2/3}\right)\right\},
\label{4}
\end{equation}
where $\gamma(a,x)$ is the lower incomplete Gamma function. In the near-boundary regime $z\ll1$, where $\tau\simeq z$, this reduces to
\begin{equation}
\phi(z)\propto
\left(\frac{z}{\tilde{u}}\right)^{-2/3},
\label{5}
\end{equation}
recovering the asymptotic scaling of the positional boundary layer (see Appendices~A and ~C for details). 

A stringent test of the proposed theory that the positional distribution can be reconstructed from orientational fluctuations is its ability to reproduce the positional distributions obtained from simulations. As demonstrated in Fig.~\ref{fig2}, the solid lines from Eq.~\eqref{4} accurately fit the simulation data across the full range of confinement strengths, capturing both the near-boundary algebraic scaling with exponent $-2/3$ and the crossover toward the interior controlled by the exponential factor. The collapse of the positional distributions in the inset of Fig.~\ref{fig2} under the rescaling transformation $z\rightarrow z/\tilde{u}$ confirms that the confinement strength  parameter $\tilde{u}$ governs the universal scaling behavior of the positional boundary layer.

In summary, the orientational dynamics of an ABP near a boundary is characterized by two key features: the emergence of stable tangential states that sustain boundary-sliding, and the orientational fluctuations around these states quantified by the conditional variance, which together govern the positional distribution within the confinement. More broadly, any quantitative characterization of ABP motion inside a circular confinement, including transport, exploration, and escape times, is fundamentally controlled by the stochastic switching dynamics between the clockwise and counterclockwise sliding states. While such problems are often studied within conventional mean first-passage time (MFPT) frameworks, these approaches do not explicitly account for the boundary-induced orientational dynamics discussed here \cite{bechinger2016active}. In particular, the accessible orientational freedom around the tangential states $\pm\pi/2$ determines the mean waiting time, associated with switching between the two sliding states. We quantify the switching dynamics in the following section.

\section{Flipping Dynamics}\label{sec-flipping}
\begin{figure}[htbp]
    \centering \includegraphics[width=1.0\columnwidth]{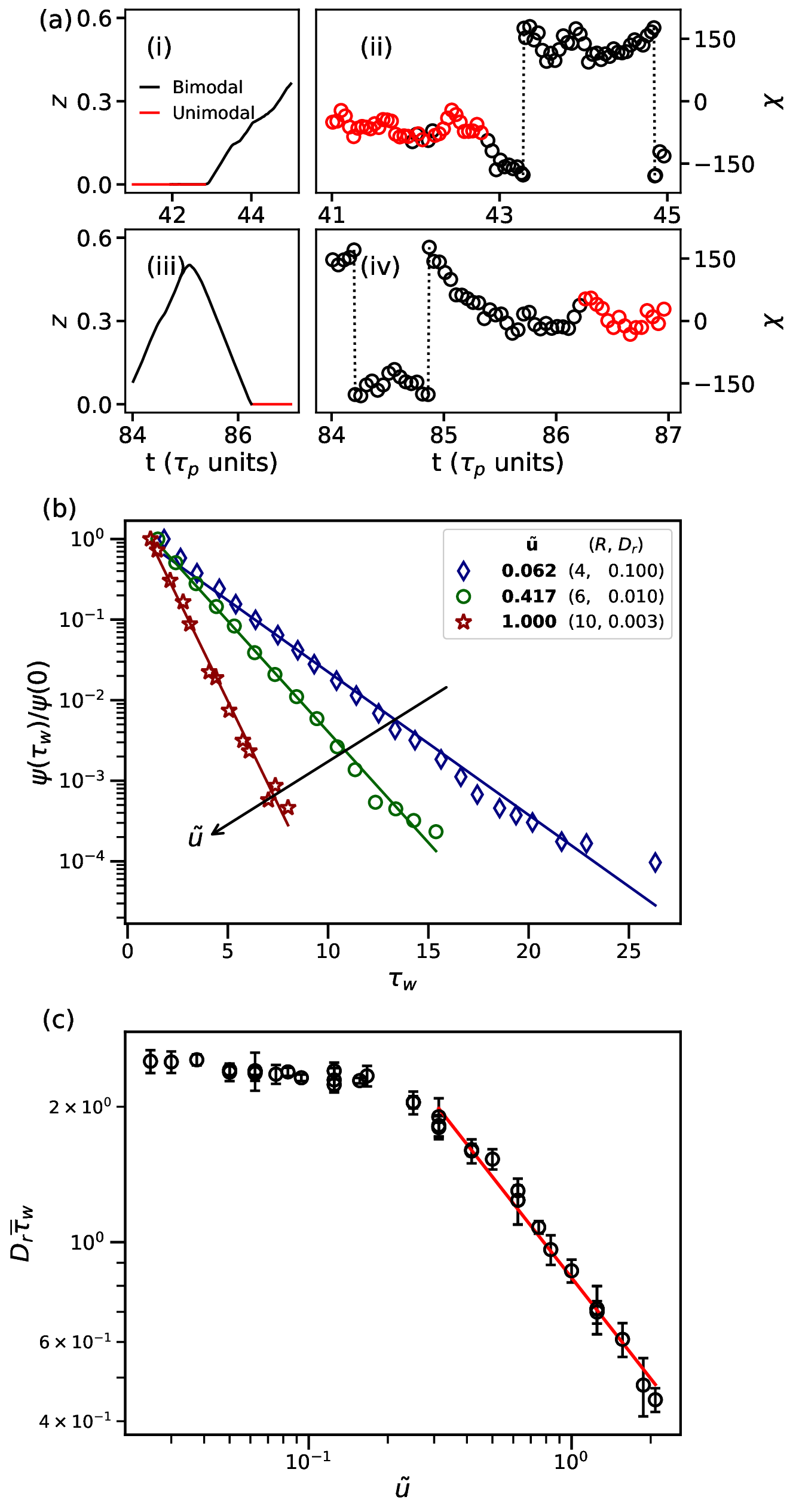}
	\caption{{\bf Flipping Dynamics} (a) Representative trajectory from simulations showing the scaled position $z$ and relative orientation $\chi$ over a selected time window. Panels (i) and (iii) show the corresponding positional dynamics, while panels (ii) and (iv) highlight fast and slow flipping pathways between the two tangential orientational states within the bimodal regime of $F(\chi|z)$. Dotted lines indicate sign reversals of $\chi$ between clockwise and anticlockwise boundary-sliding states. (b) Waiting-time distributions $\psi(\tau_w)/\psi(0)$ for different confinement strengths $\tilde{u}$, where $\psi(0)$ denotes the peak value of the normalized distribution. The time interval during which the ABP retains the same sign of $\chi$ defines the waiting time $\tau_w$. Symbols represent simulation data and solid lines denote exponential fits from which the mean waiting time $\overline{\tau}_w$ is extracted. (c) Mean waiting time scaled by the persistence time $\tau_P=1/D_r$, shown as $D_r\overline{\tau}_w$, as a function of confinement strength $\tilde{u}$. Symbols represent simulation data and the solid line corresponds to the theoretical prediction from Eq.~\eqref{eq:KRAMERS}.}
    \label{fig5}
\end{figure}

In Fig.~\ref{fig5}(a), we show a representative trajectory of confined ABP, including both particle position $z$ and orientation $\chi$ over a selected time window. Inspection of the trajectory reveals two distinct flipping pathways. Slow modes, associated with the unimodal regime of $F(\chi|z)$, occur through bulk-mediated reorientations, whereas fast modes, associated with the bimodal regime of $F(\chi|z)$, arise from boundary-localized fluctuations that switch directly between the tangential states without traversing the bulk. We characterize the orientational dynamics through stochastic switching between these tangential states, where a change in the sign of $\chi$ corresponds to the switching between clockwise and anticlockwise motion along the boundary. Although displacement of conditional peaks away from $\pm\pi/2$ typically favors slow bulk-mediated pathways, rare fast transitions remain accessible through boundary-localized fluctuations. The time interval during which the ABP retains the same sign of $\chi$ defines the waiting time $\tau_w$. For each confinement strength, we perform simulations over millions of time-steps to collect sufficient statistics and construct the corresponding waiting-time distribution $\psi(\tau_w)$, shown in Fig.~\ref{fig5}(b). The distributions exhibit an exponential form, indicating effective Poissonian like switching dynamics between the two orientational states. Adapting $\psi(\tau_w)$ to an exponential function, we extract the mean waiting time $\overline{\tau}_w$. Increasing the confinement strength $\tilde{u}$ leads to progressively steeper distributions and therefore shorter mean waiting times, corresponding to an increased flipping rate. The scaled mean waiting time $D_r \overline{\tau}_w$ obtained from simulations as a function of the confinement strength parameter $\tilde{u}$ is shown in Fig.~\ref{fig5}(c). At low confinement strengths, the mean waiting time approaches a plateau since the orientational dynamics are weakly influenced by the boundary, leading to nearly unconstrained rotational diffusion. In contrast, increasing $\tilde{u}$ enhances confinement-induced reorientation, resulting in a systematic reduction of $\overline{\tau}_w$ that follows a power-law scaling at larger confinement strengths. 

\begin{figure}[htbp]
    \centering  \includegraphics[width=1.0\columnwidth]{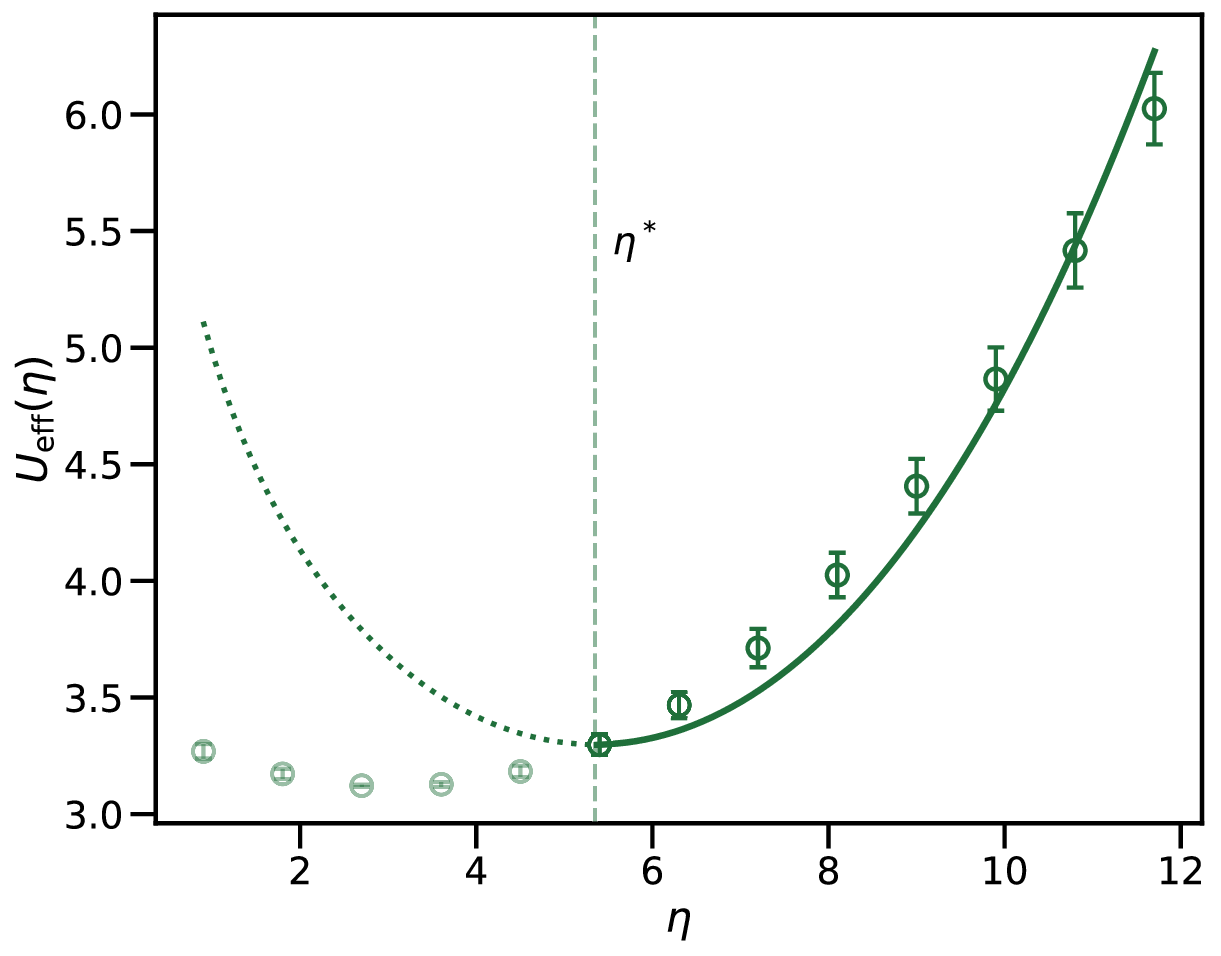}  
	 \caption{\textbf{Flipping Dynamics:} Effective potential $U_{\rm eff}(\eta)$ as a function of $\eta$, obtained from $U_{\rm eff}(\eta)=-\ln g(\eta|\tau)$, for a representative case, with $\tilde{u}=1$ at $z=2\times10^{-4}$. Symbols represent simulation data and lines denote the theoretical prediction from Eq.~\eqref{eq:EFFPOTENTIAL}. The dashed line marks the potential minimum $\eta^*$.}
 
    \label{figA3}
\end{figure}

To understand the observed switching dynamics, we construct an effective Kramers description \cite{kramers1940brownian} in terms of escape in an orientational landscape. Using the conditional orientational distribution, an effective bistable landscape may be obtained, with minima corresponding to the two preferred tangential states (see Fig.~\ref{figA3}). Within this picture, flipping between clockwise and anticlockwise motion may be interpreted as transitions between the two orientational states.
This effective escape framework predicts a power-law dependence of the mean waiting time on the confinement strength,
\begin{equation}
D_r\overline \tau_w\propto\tilde u^{-3/4},
\label{eq:KRAMERS}
\end{equation}
establishing a direct connection between confinement-induced orientational fluctuations and the stochastic switching dynamics. The theoretical prediction in Eq.~\eqref{eq:KRAMERS} is in good agreement with the simulation data shown in Fig.~\ref{fig5}(c), where the same power-law scaling is observed over the large-$\tilde u$ regime (see Appendix~D for details).

\section{Conclusions}
In this work, we have investigated the dynamics of an active Brownian particle confined within a hard circular boundary using molecular simulations and the Fokker–Planck equation. In the high-Péclet limit with vanishing translational diffusion, we showed that the positional distribution is directly coupled to orientational fluctuations, through the conditional orientational distribution. The positional distribution exhibits a power-law decay with distance from the boundary before crossing over to a uniform bulk profile. The corresponding conditional orientational distribution develops curvature-induced bistable tangential states associated with persistent boundary sliding, while the variance around these states governs the stochastic flipping between clockwise and anticlockwise sliding states. In particular, we identified two distinct flipping pathways: rapid boundary-localized flips and slower bulk-mediated excursions. The mean waiting time between flips exhibits power-law scaling with confinement strength, establishing stochastic orientational switching as a key dynamical mechanism underlying confined active motion. 

Confinement and asymmetric geometries are known to steer, trap, and sort active particles by exploiting persistent boundary-following dynamics and chirality-dependent motion \cite{bechinger2016active}. Funnel barriers \cite{drocco2012bidirectional}, ratchet geometries \cite{volpe2011microswimmers}, and Tesla-valve configurations \cite{Rogers2023} have all been shown to generate directed transport and motility-dependent sorting through repeated particle-boundary interactions. Similarly, confinement-induced boundary accumulation strongly influences exploration and escape dynamics in confined active systems \cite{militaru2021escape, aranson2022confinement, Kumar2019}. The present work suggests that such transport, trapping, and escape processes may be fundamentally governed by the emergence of bistable tangential states and distinct stochastic flipping pathways, which provide a possible microscopic mechanism to control directional persistence, reversals, boundary residence times, and escape trajectories in confined active environments. Extending these ideas to three-dimensional confinements, particularly spherical \cite{Fily2016, Nikola2016} and cylindrical geometries \cite{Morin2018}, may further clarify the role of curvature and orientational switching in steering active transport and escape dynamics in complex confined environments. It would also be interesting to explore how these confinement-induced bistable states and stochastic switching dynamics are modified, enhanced, or altered in the presence of hydrodynamic interactions known to strongly influence active motion near boundaries \cite{berke2008hydrodynamic, Brotto2013}.

Recent studies suggest that confinement not only restricts and steers the motion of active particles, but can also be influenced by the particles themselves through feedback between active dynamics and the surrounding environment, thereby enabling controlled self-organization and tunable modification of the background medium \cite{araujo2023steering}. One important manifestation of such environmental influence arises through the mechanical stresses, effective interactions, and active forces generated by self-propelled particles \cite{bechinger2016active, Caprini2019}. In particular, strongly confined self-propelled particles generate persistent active pressure on confining boundaries, with the pressure strongly dependent on the confinement size and geometry \cite{fily2014dynamics}. More recently, active particles have been used as dopants to tune yielding, relaxation, and memory effects in amorphous solids and dense passive media \cite{sharma_NPhys_2025, priya2025inverse}. From this perspective, the stochastic orientational switching dynamics identified here may additionally influence stress transmission, force fluctuations, and active-force generation in confined active-passive systems. Extending the present framework to heterogeneous and complex confinements, including rough, disordered, or dynamically evolving boundaries, may therefore provide further insight into transport, mechanical response, and emergent organization in complex active environments.

\section*{Appendix A}

%\subsection{Some useful exact results}
% The expression in Eq.~\ref{eq:AIRY} is limited to $z\ll 1$. In order to extend our analysis to the complete range of $z$, a few exact relations are derived in this subsection. 

To analyze how the stationary distribution evolves away from the boundary, we introduce the time-like variable $\tau=-\ln(1-z)$, which maps the radial distance from the wall to an effective, time-like evolution parameter. In this representation, the boundary $z=0$ corresponds to the ``initial'' value $\tau=0$, while the center of the domain ($z=1$) maps to $\tau=\infty$. The joint probability density and positional distribution in the new variable are defined as 
\begin{eqnarray}
P_1(\tau,\chi)=e^{-2\tau}P(1-e^{-\tau},\chi)\nonumber\\
\phi_1(\tau)=e^{-2\tau}\phi(1-e^{-\tau})
\label{t5+}
\end{eqnarray}
so that the new distribution is normalized as $\int_{0}^{\infty}d\tau\int_{-\pi}^{\pi}d\chi P_1(\tau,\chi)=1$. Let us now define the functions 
\begin{equation}
\lambda_n(\tau)=\int_{-\pi}^{\pi}P_1(\tau,\chi)\cos^n\chi d\chi~~~~(n=1,2,...)
\label{eq:lambda}
\end{equation}
which can be used to define the conditional moments of $\cos\chi$: 
\begin{equation}
\lambda_n(\tau)=\phi_1(\tau)\langle \cos^n\chi\rangle_\tau.
\label{eq:cond-moment}
\end{equation}
Now, consider Eq~\ref{eq:FPE-f} in the stationary case. Upon integration over $\chi$, we find that the second term on the r.h.s vanishes identically, leading to the condition 
\begin{equation}
\frac{\partial}{\partial z}\int j_z(z,\chi)d\chi=0
\end{equation}
which implies $\int j_z d\chi$ is independent of $z$. Due to the reflecting boundary condition, the integrand is identically zero at the boundary, hence $\int j_z(z,\chi)d\chi=0$ everywhere. From the expression in Eq.~\ref{eq:current}, this implies that $\lambda_1(\tau)=0$ and therefore $\langle \cos\chi\rangle_\tau=0$ for $\tau>0$. Numerical simulation results are in complete agreement with this prediction (see top inset of Fig.~\ref{fig3}). 

%\begin{figure}[htbp]
%    \centering  \includegraphics[width=1.0\columnwidth]{ApFig1.eps}  
%	\caption{ Conditional mean of $\cos\chi$ as functions of $z$ for different $(R,D_r)$ parameter sets.
% }

 %   \label{figA1}
%\end{figure}

% \begin{equation}
% \frac{\partial \lambda_1}{\partial \xi}=-\frac{\lambda_1}{\xi}
% \label{eq:lambda1}
% \end{equation}
Next, we multiply Eq.~\ref{eq:FPE-f} by $\cos\chi$ and integrate over $\chi$, leading to the equation 
\begin{equation}
\frac{d\lambda_2}{d\tau}=2\lambda_2-\phi_1.
\label{eq:lambda2}
\end{equation}
Since $\langle \cos\chi\rangle=0$, let us write $\lambda_2(\tau)=\phi_1(\tau)\sigma^2(\tau)$ where $\sigma^2(\tau)=\langle\langle \cos^2\chi\rangle\rangle_{\tau}$ is the conditional variance of $\cos\chi$. Substituting this relation into Eq.~\ref{eq:lambda2} and solving it gives 
\begin{equation}
\phi_1(\tau)\propto \frac{1}{\sigma^2(\tau)}e^{2\tau-\int_{0}^{\tau}\frac{d\tau^{\prime}}{\sigma^2(\tau^{\prime})}}
\label{eq:lambda2-phi}
\end{equation}
which leads to Eq.~\ref{eq:lambda2-phi*} after the replacement $\tau\to z$. 

Consider the limit $\tau\to 0$ in Eq.~\ref{eq:lambda2-phi}. In this case, the exponential factor tends to unity, which leads to $\sigma^2(\tau)\propto \phi_1^{-1}$, which when combined with Eq.~\ref{eq:GAMMA-TAU3} leads to the asymptotic expression 
\begin{equation}
%\sigma^2(\tau)\sim \frac{1}{\nu c}\left(\frac{\tau}{\tilde u}\right)^{\frac{2}{3}}~~~~\tau\to 0
\sigma^2(\tau)\propto \left(\frac{\tau}{\tilde u}\right)^{\frac{2}{3}}~~~~\tau\to 0
\label{eq:sigma2}
\end{equation}

Eq.~\ref{eq:sigma2} may be used to obtain a more complete understanding of the function $\sigma^2(\tau)$. We propose the scaling form 
\begin{equation}
\sigma^2(\tau)\simeq \frac{1}{2}\lambda\left(\frac{\tau}{\tilde u}\right)
%\sigma^2(\tau) \simeq \frac{1}{2}\bigg(1-e^{-(\beta\tau)^{\frac{2}{3}}}\bigg)
\label{eq:sigma22}
\end{equation}
where the scaling function $\lambda(x)$ is postulated to have the following asymptotic behaviour at small and large $x$:
\begin{eqnarray}
\lambda(x)&\propto& x^{\frac{2}{3}}~~~~~~~x\to 0\\
\lambda(x)&=&1 ~~~~~~~~~ {x\to\infty}.
\label{eq:lambda-3}
\end{eqnarray}
 The second limit ensures that the correct asymptotic behaviour is predicted at the centre of the well, i.e., $\sigma^2(\tau)\to 1/2$ as $\tau\to\infty$ corresponding to an isotropic angular distribution. 

Our numerical simulations (see Fig.~\ref{fig4}) indicate that the scaling function in Eq.~\ref{eq:lambda-3} is well-approximated by $\lambda(x)= 1-\exp(-x^{\frac{2}{3}})$, which leads to the expression in Eq.~\ref{8} for $\tau\ll 1$.
 
%  The constant $b$ can be fixed by comparison between Eq.~\ref{eq:lambda} and Eq.~\ref{eq:sigma2}, which predicts that 
% \begin{equation}
% b=\left(\frac{2}{\nu c}\right)^{\frac{3}{2}}. 
% \label{eq:beta}
% \end{equation}

\section*{Appendix B}

 The symmetry $\chi\to -\chi$ in the distribution implies that it is sufficient to focus on $\chi\geq 0$: also, the boundary condition in Eq.~\eqref{t5} suggests that, close to $\tau=0$, it is convenient to define the angular deviation $\eta=\chi-\pi/2$ for $\chi>0$. For small $\eta$, we have $\cos\chi=-\sin\eta\simeq \eta$ and $\sin\chi\simeq 1$. After denoting the transformed probability density by $\Phi_1(\tau,\eta)\equiv P_{1}(\tau,\pi/2+\eta)$, the Fokker–Planck equation in Eq.~\eqref{t2} takes the form 
\begin{equation}
\eta\frac{\partial \Phi_1}{\partial \tau}=\frac{e^{-\tau}}{{\tilde u}}\frac{\partial^2 \Phi_1}{\partial \eta^2}+\frac{\partial \Phi_1}{\partial \eta},
\label{eq6}
\end{equation}
which resembles a time-dependent Fokker-Planck equation with a time-dependent diffusion coefficient. In this representation, the spatial evolution of the probability density as one moves radially inward from the boundary is thus approximately mapped onto the temporal evolution of a diffusion process with an exponentially decaying diffusion coefficient.

%\subsection*{Partial solution using Laplace transforms}
To solve Eq.~\ref{eq6}, we use Laplace transforms. Define ${\tilde \Phi_1}(s,\eta)=\int_{0}^{\infty}e^{-s\tau}\Phi_1(\tau,\eta)d\tau$, which satisfies the equation below  for $\eta\neq 0$: 
\begin{equation}
\frac{1}{\tilde{u}}\frac{\partial^2}{\partial \eta^2}{\tilde \Phi_1}(s+1,\eta)+\bigg(\frac{\partial}{\partial\eta}-s\eta\bigg) {\tilde \Phi_1}(s,\eta)=0
\label{eq:FPE-laplace}
\end{equation}
where we have used the boundary condition in Eq.~\ref{t5}. In order to bring Eq.~\ref{eq:FPE-laplace} to a standard form, let us apply a scaling transformation and define the variable $\eta^{\prime}=\eta({\tilde u}s)^{\alpha}$ where $\alpha>0$, along with the corresponding distribution 
\begin{equation}
\Psi(s,\eta^{\prime})=({\tilde u}s)^{\alpha}{\tilde \Phi_1}\bigg(\eta^{\prime}({\tilde u}s)^{-\alpha},\chi\bigg)
\label{eq:Psi-tau}
\end{equation}
which obeys the equation
\begin{equation}
\frac{\partial^2}{\partial \eta^{\prime 2}}\Psi(s+1,\eta^{\prime})+\bigg({\tilde u}({\tilde u}s)^{-\alpha}\frac{\partial}{\partial \eta^{\prime}}-({\tilde u}s)^{1-3\alpha}\eta^{\prime}\bigg){\tilde \Psi}(s,\eta^{\prime})=0.
\label{eq:FPE-laplace2}
\end{equation}
Choosing $\alpha=1/3$ simplifies Eq.~\ref{eq:FPE-laplace2} to 
\begin{equation}
\frac{\partial^2}{\partial \eta^{\prime 2}}\Psi(s+1,\eta^{\prime})-\eta^{\prime}\Psi=-{\tilde u}({\tilde u}s)^{-\frac{1}{3}}\frac{\partial \Psi}{\partial \eta^{\prime}}. 
\label{eq:FPE-laplace3}
\end{equation}
Let us now focus on the regime $\tau\ll 1$, which maps to $s\gg 1$ in Laplace space. In the original variables, this is equivalent to considering the region close to the boundary ($r\simeq R)$. In this limit, we may approximate $\Psi(s+1,\chi)\simeq \Psi(s,\chi)$ in Eq.~\ref{eq:FPE-laplace3}. Also, for ${\tilde u}\geq 1$, we may also ignore the r.h.s in this regime, whence Eq.~\ref{eq:FPE-laplace3} reduces to the standard homogeneous Airy equation \cite{abramowitz1948handbook}, the solution for which is 
\begin{equation}
\Psi(s,\eta^{\prime})=\Gamma(s){\rm Ai}(\eta^{\prime})
\label{eq:AIRY}
\end{equation}
where the prefactor $\Gamma(s)$ is, in general, an undetermined function of the Laplace frequency $s$. The Airy function shows decaying behaviour for $\eta^{\prime}>0$ and damped oscillatory behaviour for $\eta^{\prime}<0$. However, we shall assume henceforth that $\Psi(s,\eta^{\prime})=0$ for $\eta^{\prime}<0$ %(see the discussion following Eq.~\ref{eq:NEW}). 

After using the expressions in Eq.~\ref{eq:Psi-tau} in Eq.~\ref{eq:AIRY}, it follows that 
\begin{equation}
{\tilde \phi}_1(s)=A\Gamma(s)
\label{eq:GAMMA1}
\end{equation}
where ${\tilde \phi}_1(s)$ is the Laplace transform of $\phi_1(\tau)$ and
\begin{equation}
A=2\int_{0}^{\infty} {\rm Ai}(x)dx
\label{Airy-A}
\end{equation}
is a dimensionless constant. The lower limit of integration is put to zero as already discussed. 

% In Appendix C, we prove the following exact relation between the conditional variance of $\cos\chi$ and the positional distribution function:

% \begin{equation}
% \phi(\tau)=\frac{\lambda_2(0)}{\sigma^2(\tau)}e^{2\tau-\int_{0}^{\tau}\frac{d\tau^{\prime}}{\sigma^2(\tau^{\prime})}}, 
% \label{eq:lambda2-phi}
% \end{equation}
% which predicts that the spatial dependence of the probability density $\phi(\tau)$ is controlled by orientational fluctuations. 
\section*{Appendix C}

To find the function $\Gamma(s)$ in Eq.~\ref{eq:AIRY}, let us now express Eq.~\ref{eq:lambda2} in Laplace space: 
\begin{equation}
{\tilde \lambda_2}(s)(s-2)+{\tilde\phi_1}(s)=\lambda_2(0)
\label{eq:laplace4}
\end{equation}
For $\chi\simeq \pi/2$, $\cos^2\chi\simeq \eta^2$, and it therefore follows from Eq.~\ref{eq:AIRY} that 
\begin{equation}
{\tilde \lambda_2}(s)=B({\tilde u}s)^{-\frac{2}{3}}\Gamma(s)
\label{eq:GAMMAS}
\end{equation}
where 
\begin{equation}
B=2\int_{0}^{\infty}{\rm Ai}(x)x^2dx
\end{equation}
Substitute Eq.~\ref{eq:GAMMAS} in Eq.~\ref{eq:laplace4} to arrive at 
\begin{equation}
\Gamma(s)=\frac{\lambda_2(0)}{A+B(s-2)({\tilde u}s)^{-\frac{2}{3}}}
\label{eq:GAMMAS1}
\end{equation}
In the regime $\tau\ll 1$, we may assume $s\gg 2$ in Eq.~\ref{eq:GAMMAS1}, which leads to the simpler expression 
\begin{equation}
\Gamma(s)\simeq \frac{c_1}{s^{\frac{1}{3}}+c_2{\tilde u}^{\frac{2}{3}}}
\label{eq:GAMMAS2}
\end{equation}
where $c_1=\lambda_2(0){\tilde u}^{2/3}/B$ and $c_2=A/B$. Substitution of Eq.~\ref{eq:GAMMAS1} in Eq.~\ref{eq:GAMMA1} and doing Laplace-inversion leads to the scaling form
\begin{equation}
%\phi_1(\tau)=A\frac{c_1\sqrt{3}}{2\pi}\psi(a\tau)
\phi_1(\tau)=\psi(a\tau)
\label{eq:GAMMA-TAU1}
\end{equation}
where $a=c_2^3{\tilde u}^2$ and the scaling function 
% \begin{equation}
% a=c_2^3{\tilde u}^2
% \label{eq:a}
% \end{equation}
\begin{equation}
\psi(y)\propto \int_{0}^{\infty}\frac{x^{1/3}e^{-z x}}{x^{2/3}+x^{1/3}+1}dx.
\label{eq:GAMMA-TAU2}
\end{equation}
For $a\tau\ll 1$, the limiting behaviour of the above integral is given by $\psi(y)\sim \Gamma(2/3)y^{-2/3}$, which when used in Eq.~\ref{eq:GAMMA-TAU1}, leads to 
\begin{equation}
%\phi_1(\tau)\sim A\frac{c_1\sqrt{3}\Gamma(2/3)}{2\pi}\tau^{-2/3}~~~~(\tau\ll {\tilde u}^{-2})
\phi_1(\tau)\propto \tau^{-2/3}~~~~(\tau\ll {\tilde u}^{-2})
\label{eq:GAMMA-TAU3}
\end{equation}
Next, we derive an expression for the joint distribution $\Phi_1(\tau,\eta)$ for small $\eta$. The large $s$-regime is of direct interest to us; for $s\gg {\tilde u}^2$, 
$\Gamma(s)\sim c_1 s^{-1/3}$ from Eq.~\ref{eq:GAMMAS2}, which when substituted in Eq.~\ref{eq:AIRY} and after using the scaling relation in Eq.~\ref{eq:Psi-tau} leads to 
\begin{equation}
{\tilde \Phi_1}(s,\eta)\sim \frac{\lambda_2(0){\tilde u}}{B}{\rm Ai}\left(({\tilde u}s)^{\frac{1}{3}}\eta\right)~~~s\gg {\rm max}(1,{\tilde u}^2)
\label{eq:PHI-S}
\end{equation}
For large $s$ and large ${\tilde u}$, the asymptotic form of the Airy function \cite{abramowitz1948handbook} may be used, leading to our final expression in Laplace space:
\begin{equation}
{\tilde \Phi_1}(s,\eta)\sim \lambda_2(0){\tilde u}^{11/12}s^{-1/12}\eta^{-\frac{1}{4}}e^{-\frac{2}{3}\sqrt{{\tilde u}s}\eta^{3/2}}
%{\tilde \Phi_1}(s,\eta)\propto \bigg(\frac{{\tilde u}^3}{\eta s^{\frac{1}{3}}}\bigg)^{\frac{1}{4}}e^{-\frac{2}{3}\sqrt{{\tilde u}s}\eta^{3/2}}~~~(s\gg {\tilde u}^2)
\label{eq:FINAL-LAPLACE}
\end{equation}
Laplace inversion of Eq.~\ref{eq:FINAL-LAPLACE} leads to 
\begin{equation}
\Phi_1(\tau,\eta)\propto \bigg(\frac{\tau}{\tilde{u}}\bigg)^{-\frac{17}{12}}\eta^{\frac{5}{4}}e^{-\frac{{\tilde u}\eta^3}{9\tau}}~~~~(\tau\ll {\tilde u}^{-2})
\label{FINAL-TAU}
\end{equation}
% After integrating over $\eta$, we find that 
% \begin{equation}
% \phi(\tau)\propto \tau^{-2/3}~~~~~\tau\ll {\rm min}(1,{\tilde u}^{-2})
% \label{eq:phitau}
% \end{equation}
The conditional angular distribution is defined through the relation 
\begin{equation}
g(\eta|\tau)= \frac{\Phi_1(\tau,\eta)}{\phi_1(\tau)}. 
\label{eq:CONDITIONAL1}
\end{equation}
In the limit $\tau\to\infty$, we expect the angular distribution to become uniform, i.e., $\lim\limits_{\tau\to\infty} g(\chi|\tau)=1/2\pi$, corresponding to a perfectly isotropic distribution of particle orientations. In the opposite limit, close to $\tau=0$ where the distribution is bimodal, we find, after substitutions from Eq.~\ref{eq:FINAL-LAPLACE} and Eq.~\ref{eq:GAMMA-TAU3} that, 
\begin{equation}
g(\eta|\tau)\propto \tau^{-\frac{3}{4}}\eta^{\frac{5}{4}}\exp\left(-\frac{{\tilde u}\eta^3}{9\tau}\right). 
\label{eq:CONDITIONAL2}
\end{equation}
The prefactor can be computed using normalisation: $2\int_{0}^{\infty}g(\eta|\tau)d\eta=1$. Eq.~\ref{eq:CONDITIONAL2} directly leads to the expression in Eq.~\ref{7} for the conditional angular distribution, expressed in terms of $\chi$ and $z$ (when $z\ll 1$). 

It is also easily seen that, for $\eta>0$, the expression in Eq.~\ref{eq:CONDITIONAL2} has a maximum at $\eta_{\rm max}\propto \tau^{\frac{1}{3}}$. When expressed in terms of $\chi$, this leads to the prediction in Eq.~\ref{11} in the limit $\tau\ll 1$, where $\tau\simeq z$. 

\section*{Appendix D}
Here, we show that the non-trivial scaling behaviour of the mean flip time can be explained using an effective Kramers' theory \cite{kramers1940brownian} for escape of a trapped Brownian particle over a potential barrier. The effective potential energy $U_{\rm eff}(\eta;\tau)$ to trap the particle in the angular potential wells is given by equating 
$g(\eta|\tau)= e^{-U_{\rm eff}}$, where $g(\eta|\tau)$ is the conditional angular distribution (Eq.~\ref{eq:CONDITIONAL2}). The explicit expression turns out to be 
\begin{equation}
U_{\rm eff}(\eta;\tau)=-\frac{5}{4}\log \eta+\frac{{\tilde u}\eta^3}{9\tau}
\label{eq:EFFPOTENTIAL}
\end{equation}
which has a single minimum at the point $\eta^*=(15\tau/4{\tilde u})^{1/3}$. 

The mean time of escape for the particle from such a minimum \cite{vanKampen1992} may be identified as the mean waiting time between flips in the slow regime, and is given by 
\begin{equation}
D_r{\overline \tau_w}\propto \frac{e^{U_{\rm eff}(\eta^*)}}{U_{\rm eff}^{\prime\prime}(\eta^*)}.
\label{eq:kramers}
\end{equation}
After using the explicit expression for $\eta^*$ in Eq.~\ref{eq:kramers}, we find that 
\begin{equation}
D_r{\overline \tau_w}\propto \bigg(\frac{\tau}{\tilde u}\bigg)^{1/4}
\label{eq:escapetime}
\end{equation}
In the bimodal regime, it may be assumed that most escape events occur slightly away from the boundary, where the effective potential well is shallowest. For ${\tilde u}>1$, we can identify this regime as $\tau\sim {\tilde u}^{-2}$ from Eq.~\ref{FINAL-TAU}, which approximately corresponds to the near boundary of the bimodal regime. After using this estimate in Eq.~\ref{eq:escapetime}, we arrive at the expression in Eq.~\ref{eq:KRAMERS}, which agrees well with simulation data.
% \section*{Appendix E}
% \subsection*{Global Orientational Distribution}
% \begin{figure}[htbp]
%     \centering  \includegraphics[width=0.9\columnwidth]{ApFig2.eps}  
% 	\caption{ Global orientational distribution, $G(\chi)$ for different confinement strength, $\tilde{u}$
%  }\label{figA2}
% \end{figure}
% While the orientational statistics in the bimodal regime exhibit two locally preferred orientations, the global orientational distribution obtained by integrating $F(\chi|z)$ over all $z$ is unimodal. This arises because the unimodal region near the boundary, where orientations are concentrated around $\chi = 0$, dominates the overall weighting of configurations, whereas the bimodal contributions at larger $z$ are comparatively weaker when averaged over the entire confinement. Consequently, the global distribution collapses into a single peak centered at $\chi = 0^\circ$. The resulting distribution is well captured by the Gaussian form (THEORETICAL ARGUMENT?) 
% \begin{eqnarray}
%     G(\chi )\propto \exp\!\left(-\tfrac{1}{2}\tilde{u}\chi ^{2}\right)
% \end{eqnarray}
% indicating that the confinement strength $\tilde{u}$ controls not only the local orientational structure but also the global orientational profile.

%\clearpage

\bibliography{reference}
\end{document}